\documentclass[journal=jacsat,manuscript=letter]{achemso}

\usepackage{xcolor}
\usepackage[utf8]{inputenc}
\usepackage{amsmath}
\usepackage{amssymb}
\usepackage{graphicx}
\usepackage{gensymb}
\usepackage{graphics}
\usepackage[T1]{fontenc}
\usepackage{soul}

\def\beq {\begin{equation}}
\def\eeq {\end{equation}}

\def\bfk {\mathbf{k}}
\def\bfq {\mathbf{q}}

\makeatletter

\author{Giovanni Marini}
\email{giovanni.marini-2@unitn.it}
\affiliation{Department of Physics, University of Trento, Via Sommarive 14, 38123 Povo, Italy}

\author{Matteo Calandra} 
\affiliation{Department of Physics, University of Trento, Via Sommarive 14, 38123 Povo, Italy}

\author{Pierluigi Cudazzo}
\affiliation{Department of Physics, University of Trento, Via Sommarive 14, 38123 Povo, Italy}
\alsoaffiliation{European Theoretical Spectroscopy Facility (ETSF)}

\title{Optical absorption and photoluminescence of single layer boron nitride from a first principles cumulant approach}


\begin{document}

\begin{abstract}

The photoluminescence spectrum of a single-layer boron nitride remains elusive, marked by enigmatic satellites that hint at a significant but unidentified exciton-phonon coupling.
 Here, by employing a first principles approach based on the many-body cumulant expansion of the charge response, we calculate the optical absorption and photoluminescence  of a  single layer boron nitride. 
 We identify the specific exciton-phonon scattering channels and unravel their impact on the optical absorption and photoluminescence spectra thereby providing an interpretation of the experimental features.
 Finally, we show that, even in a strongly polar material such as h-BN monolayer, the electron-hole interaction responsible for the excitonic effect results in the cancellation of the Fr\"olich interaction at small phonon momenta.  This effect is captured only if the invariance of the exciton-phonon matrix elements under unitary transformations in the Bloch functions manifold is preserved in the calculation.  
\end{abstract}


Twenty years after the synthesis of graphene\cite{doi:10.1126/science.1102896} the study of two-dimensional (2D) materials still attracts a great deal of interest, due to their peculiar properties. Among these, some insulating compounds are especially interesting due to the indirect-to-direct band gap transition, when going to the monolayer limit\cite{doi:10.1021/nl903868w}, and the presence of strong excitonic effects that open new paths in both fundamental physics and technological applications. Among 2D semiconductors, the strongly polar monolayer Boron Nitride (m-BN), with a huge exciton binding energy ($\approx 2$ eV) and a simple electronic structure, constitutes a reference system for the study of excitonic physics in two dimensions\cite{cudazzo2016,galvani2016}. In addition, m-BN is a promising candidate material for a broad range of applications in optics, electro-optics and quantum optics\cite{caldwell2019,gil2020}. For example, it hosts defects that can be engineered to obtain room-temperature, single-photon emission in the ultraviolet, visible, and near-infrared ranges, and it exhibits exceptional properties in the deep-ultraviolet for a new generation of emitters and detectors\cite{watanabe2009}.

Despite its  simple structure, m-BN is far from being well understood. Indeed, recent photoluminescence (PL) experiments performed using different substrates have shown controversial results. In particular the PL spectrum of m-BN on graphite\cite{Elias2019} displays a double peak structure arising from phonon satellites below the direct exciton energy, suggesting that m-BN is an indirect band gap semiconductor. On the other the PL spectrum of m-BN exfoliated on silicon oxide\cite{doi:10.1021/acs.nanolett.1c02531} displays a single dominant peak together with lower energy phononic sidebands typical of direct gap polar crystals. Both experiments highlight that the coupling to lattice vibrations plays a key role in setting the optical properties of this system.

In this context, being the PL spectra strongly dependent on the nature of the substrate, an essential step towards the understanding of the optical properties of this material is the first-principles study of the optical properties of the ideal suspended BN layer in order to disentangle  the substrate contribution. However, even if the electronic structure and excitonic effects on the optical spectra have been extensively investigated using both models\cite{galvani2016,henriques2020} and first-principles calculations\cite{cudazzo2016,prete2020,PhysRevMaterials.7.024006}, to the best of our knowledge, the coupling of the excitons with phonons is a problem that has never been addressed in isolated m-BN.

In this letter, by using first-principles many-body perturbation theory (MBPT) \cite{RevModPhys.74.601,strinati1988}, we present a full ab-initio study of the exciton-phonon (exc-ph) coupling in m-BN and its effect on the absorption and PL spectra.  For the first time, we employ the recently developed cumulant expansion\cite{PhysRevResearch.2.012032,PhysRevB.102.045136,PhysRevB.108.165101} for the two-particle correlation function ($\mathcal{L}$) combined with a full $ab$-$initio$ calculation of the electronic structure\cite{hybertsen1986,rohlfing1998} and vibrational properties\cite{RevModPhys.73.515} providing a new computational scheme for the study of neutral excitations that takes into account dynamical effects induced by the coupling with lattice vibration in a consistent way. Our implementation preserves the rotational invariance of the exc-ph matrix elements in the manyfold of the Bloch states (gauge invariance), a crucial ingredient to obtain reliable results at small phonon-momenta, see section S1 of Supplemental Information (SI), that also contains citations to Refs.\cite{aryasetiawan1996,kheifets2003,guzzo2011,lischner2013,zhou2018,verdi2017,caruso2018,zhou2015,hybertsen1986,PhysRevLett.125.107401,rohlfing1998,cannuccia2019,RevModPhys.73.515,hamnn2013,QE,QE2,PhysRevB.96.075448,sangalli2019,bruneval2008,rozzi2006,dancoff1950}.
In our approach, in analogy with the standard GW plus Bethe-Salpeter equation (GW+BSE) scheme\cite{RevModPhys.74.601}, the absorption [$I_{Abs}(\omega)$] and the PL [$I_{PL}(\omega)$] signals at a given frequency ($\omega$) are directly linked to the retarded [$\mathcal{L}^R_{\boldsymbol{\lambda}}(\omega)$] and lesser [$\mathcal{L}^<_{\boldsymbol{\lambda}}(\omega)$] components of the two particle correlation function $\mathcal{L}_{\boldsymbol{\lambda}}$  evaluated in the basis of excitonic states (being $\boldsymbol{\lambda}=(\lambda,\bfq)$ a compact notation to indicate the exciton band index $\lambda$ and the corresponding wave vector $\bfq$, see SI for more details). The optical properties are obtained by  
taking the sum of $\mathcal{L}_{\boldsymbol{\lambda}}$ over all the excitonic states with zero wave vector weighted by the corresponding exciton dipole matrix elements ($d_{\boldsymbol{\lambda}}$), so that:
$I_{Abs}(\omega)\propto\sum_{\boldsymbol{\lambda}}|d_{\boldsymbol{\lambda}}|^2\mathcal{L}^R_{\boldsymbol{\lambda}}(\omega)$\cite{strinati1988} and
$I_{PL}(\omega)\propto i\omega^2\sum_{\boldsymbol{\lambda}}|d_{\boldsymbol{\lambda}}|^2\mathcal{L}^<_{\boldsymbol{\lambda}}(\omega)$\cite{hannewald2000}.

\begin{figure}[t!]
\centering
\includegraphics[width=0.8\linewidth]{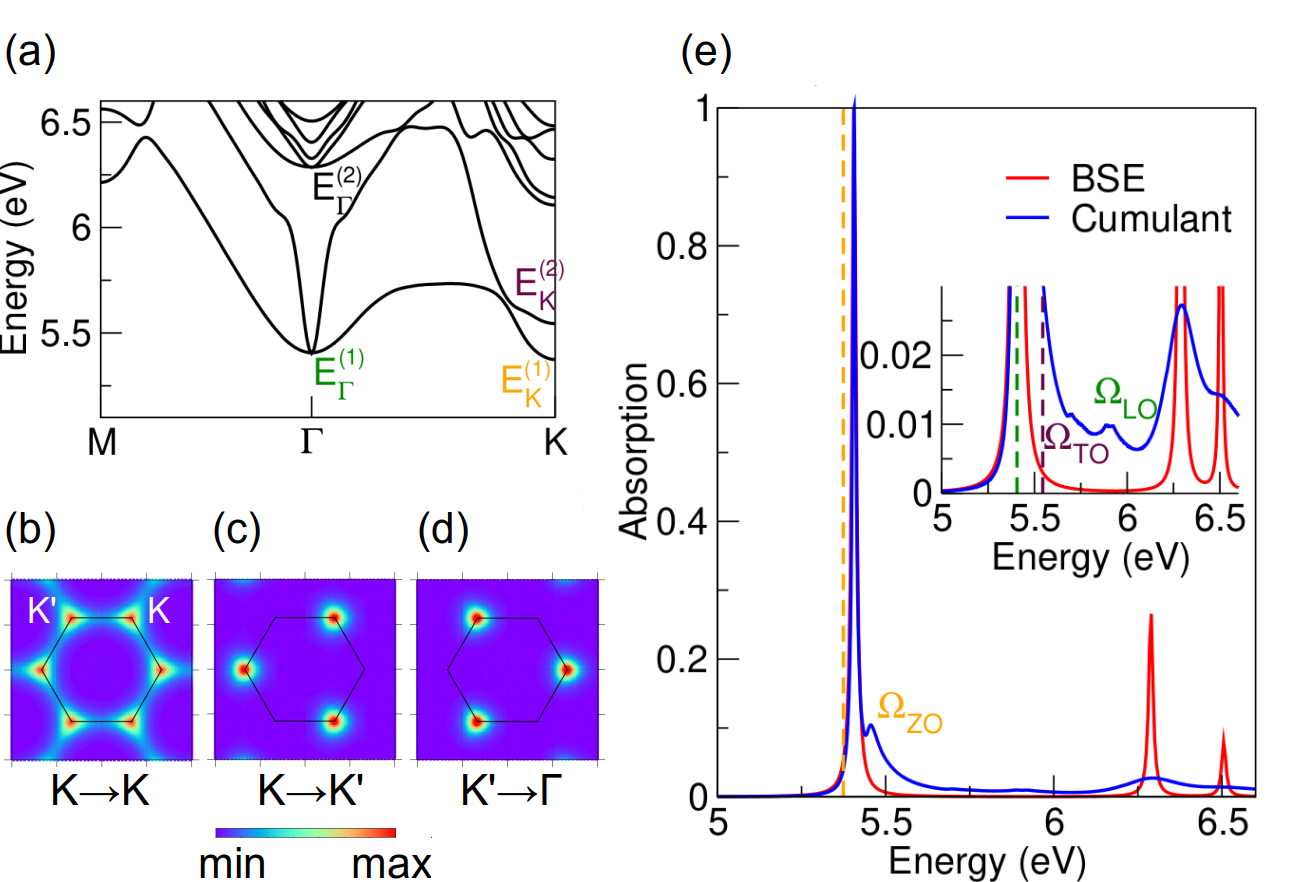}
\caption{Panel (a): excitonic band structure. Panels (b)-(d): excitonic amplitudes for the $E^{(1)}_{\Gamma}$, $E^{(2)}_{K}$ and $E^{(1)}_{K}$ excitons, respectively. The corresponding hole $\rightarrow$ electron transition is reported below each panel. Panel (e): absorption spectra evaluated in different approximations. Each spectrum is renormalized to the intensity of the highest peak.}\label{fig1}
\end{figure}

A necessary initial step for the study of optical properties is the assessment of the electronic structure. Our G$_0$W$_0$ calculation shows that the electronic band structure is noticeably modified by many-body effects. Beside a direct-gap opening of 2.8 eV at the $K$ point of the first BZ, quasi particle (QP) corrections to the Kohn-Sham band structure also induce a direct-indirect band gap transition due to a nearly free-electron conduction state at $\Gamma$ descending below the bottom of the conduction band at $K$ . Our findings are consistent with previous works\cite{prete2020,PhysRevMaterials.7.024006} [see SI for additional details]. The same behavior is observed when excitonic effects are included through the BSE: as can be seen from the excitonic band structure in Fig. \ref{fig1} (a), the energy of the indirect exciton at $\bfq=K$ ($E^{(1)}_K$) is 30 meV below the energy of the direct exciton ($E^{(1)}_{\Gamma}$). The character of this excitation can be understood looking at the corresponding excitonic amplitude $A^{cv\bfk}_{\lambda\bfq}$ that, for a given exciton in state $(\lambda,\bfq)$, defines the weight of the independent particle transition $(v,\bfk)\rightarrow(c,\bfk+\bfq)$ in the excitonic wave function ($v$ and $c$ running on valence and conduction bands, respectively). As we can see, the $\bfk$-space distribution of the holes in valence band (Fig. \ref{fig1} (d)) is localized around the $K'$ point showing unequivocally how this state originates from the superposition of electron-hole pairs involving the nearly free electron band at $\Gamma$ and the top of the valence band at $K'$. Moreover, since the same wave-vector $\bfq=K$ also connects the two nonequivalent valleys at $K$ and $K'$, an inter-valley exciton appears at higher energy ($E^{(2)}_{K}$, Fig. \ref{fig1} (a)). The corresponding excitonic amplitude is localized close to the $K$-point and its nature is the same as the one of the direct exciton (cfr. Fig. \ref{fig1} (b) and (c)), closely resembling what is observed in other valley semiconductors with hexagonal structure such as transition metal dichalcogenides\cite{MOLINASANCHEZ2015554}.

\begin{figure}[t!]
\centering
\includegraphics[width=0.8\linewidth]{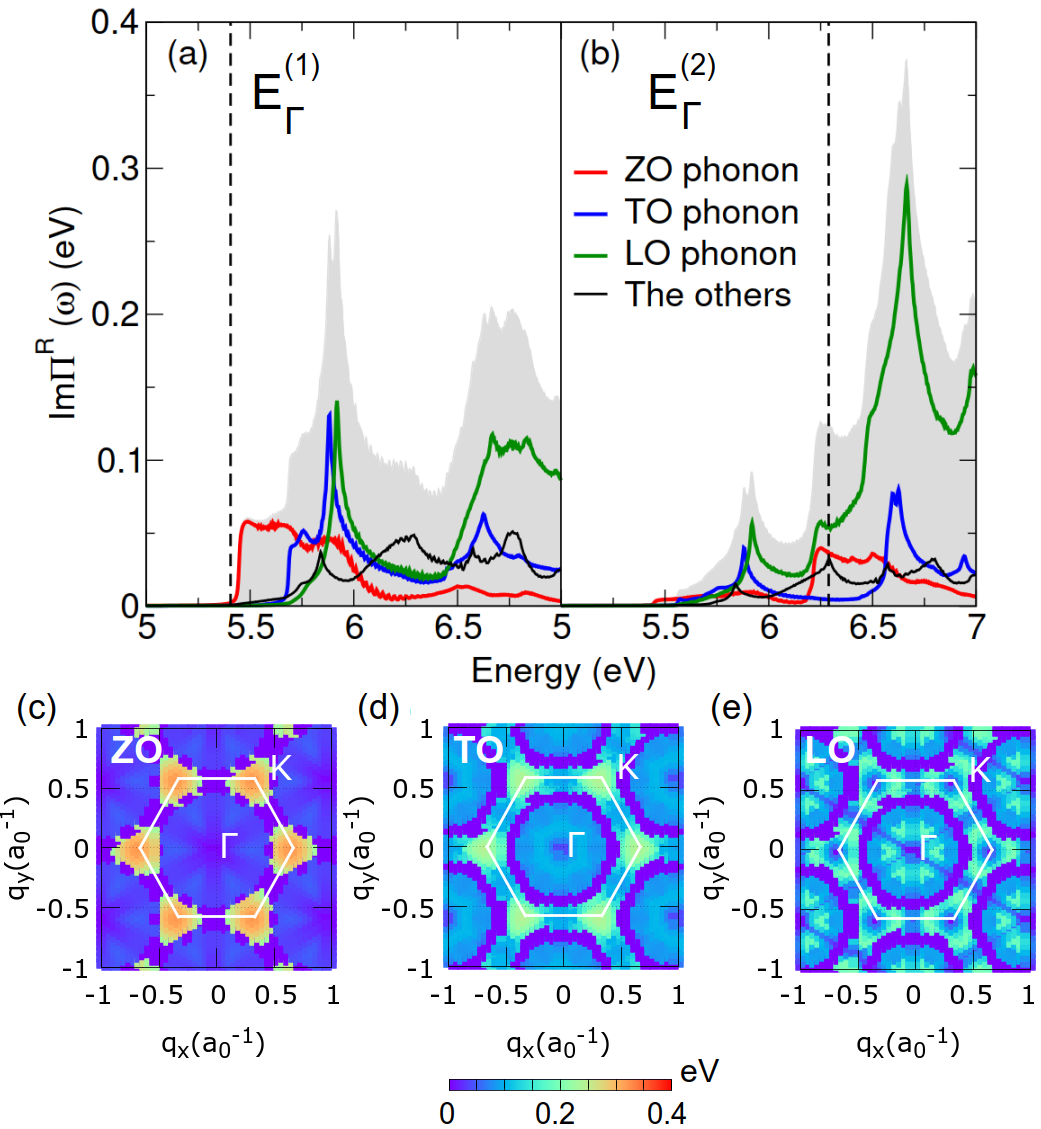}
\caption{Imaginary part of the retarded exciton self-energy (gray region) resolved in the different phononic contributions (continuous) lines for the first two direct excitons: $E^{(1)}_{\Gamma}$ (panel (a)) and $E^{(2)}_{\Gamma}$ (panel (b)). The dashed line indicates the position of the exciton. Panels (c)-(e): symmetrized exciton-phonon coupling, $\sqrt{\sum^{2}_{\lambda=1}|\mathcal{G}_{\lambda\alpha;\mu}(\bfq=0,\bar{\bfq})|^2}$ as a function of $\bar{\bfq}$ for ZO ($\alpha$=1),  TO ($\alpha$=2) and LO($\alpha$=1) modes.  }\label{fig2}
\end{figure}
In Fig. \ref{fig1} (e) we show the optical absorption spectrum obtained from the solution of the BSE (red line) using a broadening of 5 meV that matches the estimated radiative linewidth of the lowest excited state of m-BN\cite{henriques2020}. Consistently with previous calculations available in literature\cite{cudazzo2016,galvani2016,PhysRevMaterials.7.024006}, the onset of the spectrum is set by the dominant peak at 5.40 eV which is associated to the lowest direct exciton ($E^{(1)}_{\Gamma}$). The lesser pronounced peaks at 6.30 eV and 6.60 eV arise from the exciton $E^{(2)}_{\Gamma}$ and higher energy states (not shown) with the same character of the lowest excited state.  

In order to include the effect of the lattice vibrations, we performed a cumulant expansion of $\mathcal{L}^R_{\boldsymbol{\lambda}}$ in terms of the retarded exciton self-energy $\Pi_{\boldsymbol{\lambda}}^R$ evaluated at the leading order in the exc-ph matrix elements $\mathcal{G}_{\lambda\alpha;\mu}(\bfq,\bar{\bfq})$\cite{PhysRevB.102.045136,PhysRevB.108.165101,PhysRevB.105.085111,PhysRevLett.125.107401}. These quantities define the scattering amplitude from an excitonic state $(\lambda,\bfq)$ to an excitonic state $(\alpha,\bfq+\bar{\bfq})$ mediated by a phonon with wave vector $\bar{\bfq}$ and band index $\mu$. As explained in the SI, they have the following structure:
$\mathcal{G}_{\lambda\alpha;\mu}(\bfq,\bar{\bfq})=\mathcal{A}_{\lambda\alpha;\mu}(\bfq,\bar{\bfq})-\mathcal{B}_{\lambda\alpha;\mu}(\bfq,\bar{\bfq})$, with:

\begin{equation}
\begin{gathered}
    \mathcal{A}_{\alpha\lambda;\mu}(\bfq,\bar{\bfq})= \sum_{v\bar{v}c\mathbf{k}} A^{c\bar{v}\mathbf{k-\mathbf{\bar{q}}}*}_{\alpha\mathbf{q}+\mathbf{\bar{q}}} g_{v\bar{v},\mu}(\mathbf{k} -\mathbf{\bar{q}},\mathbf{\bar{q}})A_{\lambda\mathbf{q}}^{cv\mathbf{k}} \\
    \mathcal{B}_{\alpha\lambda;\mu}(\bfq,\bar{\bfq})=\sum_{c\bar{c}v\mathbf{k}}A^{\bar{c}v\mathbf{k}*}_{\alpha\mathbf{q}+\mathbf{\bar{q}}} g_{\bar{c}c,\mu}(\mathbf{k} +\mathbf{q},\mathbf{\bar{q}})A_{\lambda\mathbf{q}}^{cv\mathbf{k}},
            \label{eqg}
            \end{gathered}
\end{equation}
where $g$ denote the electron-phonon (el-ph) matrix elements between Bloch states and, $v$ ($\bar{v}$) and $c$ ($\bar{c}$) run over the valence and conduction bands, respectively.

As can be inferred from Fig. \ref{fig1}(e), the spectrum is noticeably modified when the coupling with lattice vibrations is taken into account. The cumulant calculation done a $T$=$100$~K (blue line) clearly shows how the higher energy region of the absorption spectrum is strongly damped by the exc-ph coupling, so that no clear energy separation between the two peaks is present (see inset). On the other hand, damping is nearly negligible on the first peak, leaving its linewidth unchanged. Despite that, this excitation displays a significant renormalization (40 $\%$) as a consequence of the spectral weight transfer from the QP to the phonon satellites at higher energies. They involve, beside the peak at 5.45 eV, also less pronounced features at 5.70 eV and 5.90 eV (see inset in Fig. \ref{fig1} (e)).

Being satellites directly linked to the poles of the exciton self-energy, additional insights on their structure can be gained examining the imaginary part of $\Pi^R_{\boldsymbol{\lambda}}(\omega)$.  In Fig. \ref{fig2} we show $\operatorname{Im}\Pi^R_{\boldsymbol{\lambda}}(\omega)$ (gray region) evaluated at 100 K for the first two bright excitons ($E^{(1)}_{\Gamma}$ and $E^{(2)}_{\Gamma}$) highlighting the contribution of the different phononic branches. The onset of the spectrum for the retarded exciton self-energy related to $E_\Gamma^{(1)}$ (Fig. \ref{fig2}(a)) is set by the coupling with the out-of-plane transverse optical (ZO) phonon mode (red curve) mainly involved in scattering processes between the $E^{(1)}_{\Gamma}$ exciton and the states belonging to the first excitonic band. The same scatterings also give rise to the peak at 5.50 eV which is responsible for the appearance of the first phonon satellite in Fig. \ref{fig1} (e). The high energy satellites, on the other hand, originate from the large double peak at 5.88 eV and 5.92 eV related to the longitudinal optical (LO) and the in-plane transverse optical (TO) phonon modes mainly coupled with the first and second excitonic band, respectively\cite{loto} (see S2 of SI for the convention used to define ZO, TO and LO at finite phonon momentum). The high energy region of the retarded self-energy spectrum above 6.50 eV is dominated by the coupling between the LO phonon with higher energy excitonic states and leads to negligible effects on the optical spectrum due to the large energy difference with the $E^{(1)}_{\Gamma}$ exciton.

The extremely small spectral weight of $\operatorname{Im}\Pi^R_{\boldsymbol{\lambda}}(\omega)$ at the exciton energy explains the negligible linewidth induced by the exc-ph coupling on the first bright exciton\cite{footnote}. 
The situation is quite different for the second bright exciton, in Fig. \ref{fig2} (b). Here $\operatorname{Im}\Pi^R_{\boldsymbol{\lambda}}$ presents a shoulder at the exciton energy which is responsible for the strong damping observed in the absorption spectrum. In particular the damping is induced by scattering processes with both lower energy and higher energy states mediated by the ZO and LO phonon modes, respectively.

\begin{figure}[t!]
\centering
\includegraphics[width=0.8\linewidth]{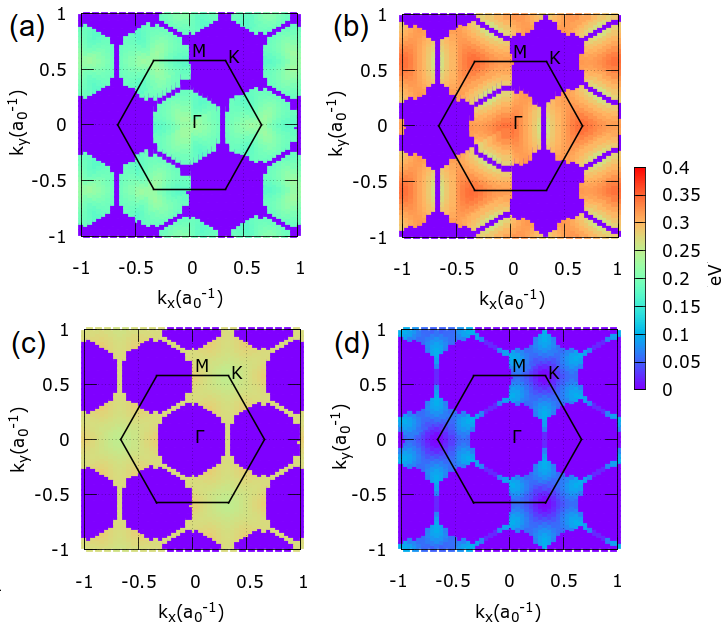}
\caption{Plot of the intraband electron-phonon coupling, $|g_{nn,\nu}(\mathbf{k},\mathbf{q}=K)|$, as a function of $\mathbf{k}$ for the top valence ($n=v$) (left column) and bottom conduction band ($n=c$) (right column). Panels (a) and (b): ZO mode ($\nu=2$). Panels (c) and (d): TO mode ($\nu=5$).}\label{fig3}
\end{figure}

To gain further insight on the nature of the exc-ph coupling in this system, we now analyze the momentum distribution of the exc-ph matrix elements relative to the $E^{(1)}_{\Gamma}$ exciton for the most relevant phonon and exciton bands. Thus, we set $\bfq=0$ in Eq.\ref{eqg} and restrict the index $\lambda$ to the two fold degenerate excitons at $\Gamma$. 
 
Moreover, based on the previous analysis carried out on the exciton retarded self-energy, we focus on the coupling of the ZO and LO phonon modes with the first exciton band ($\alpha=1$) and the coupling of the TO phonon mode with the second exciton band ($\alpha=2$).

Our results are summarized in Fig. \ref{fig2} (c)-(d) where we plot for each selected coupling channel the corresponding symmetrized quantity: $\sum^{2}_{\lambda=1}|\mathcal{G}_{\lambda\alpha;\mu}(\bfq=0,\bar{\bfq})|^2$ as a function of $\bar{\bfq}$. As we can see, the nature of the exc-ph coupling for the three phonon modes is completely different. For the ZO and TO modes (Fig. \ref{fig2} (c) and (d)) the coupling is localized around the $K$ ($K'$) point of the first BZ, highlighting the inter-valley character of the corresponding $E^{(1)}_{\Gamma}\leftrightarrow E^{(1)}_K$ and $E^{(1)}_{\Gamma}\leftrightarrow E^{(2)}_K$ scattering processes. For the LO mode, on the other hand, an intra-valley component is present as a direct consequence of the long range nature of the interaction with the LO phonons in polar crystals.

\begin{figure}[t!]
\includegraphics[width=0.8\linewidth]{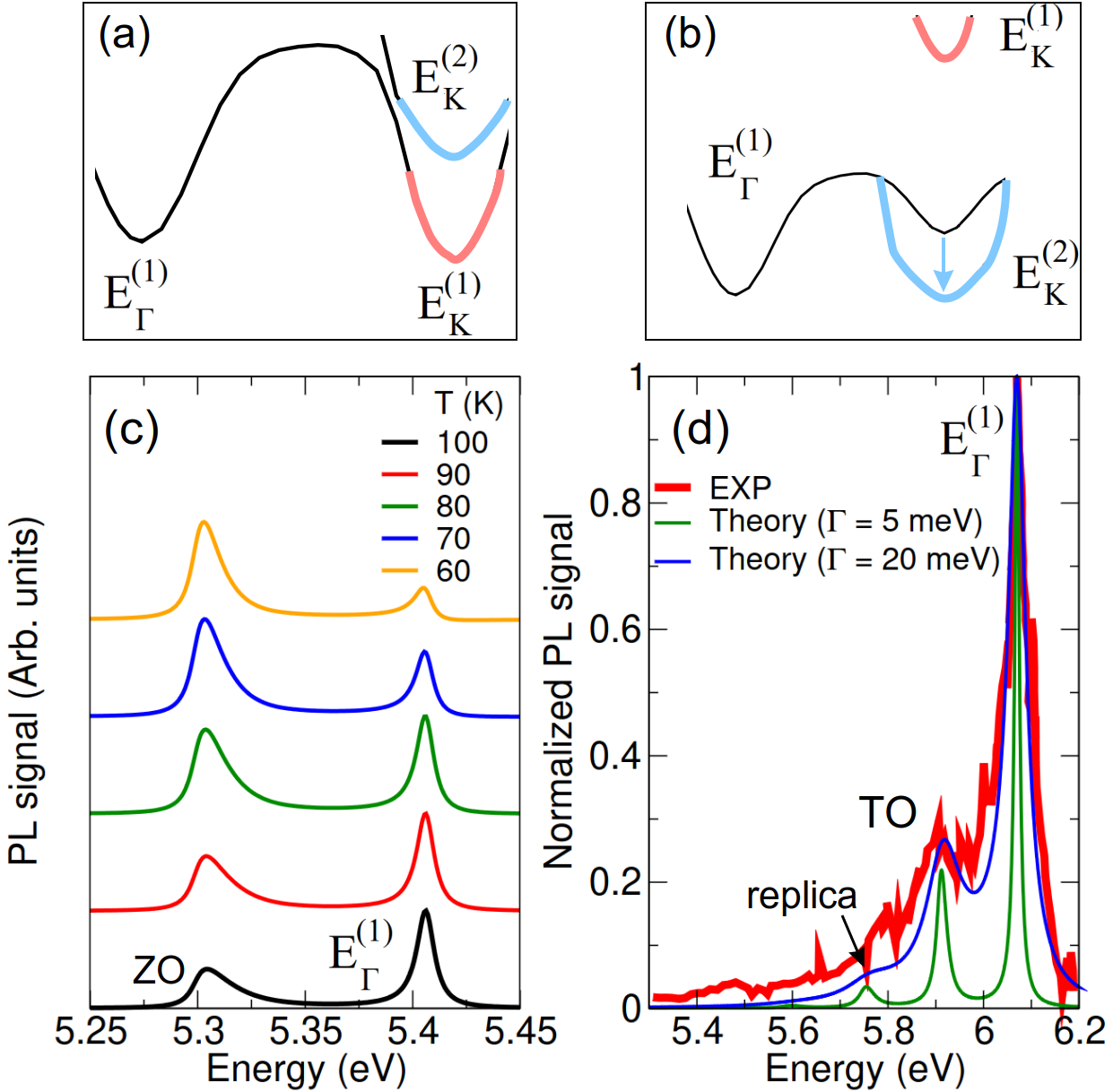}
\caption{Panel (a): schematic representation of the excitonic dispersion for of ideal m-BN. Panel (b): schematic representation of the excitonic dispersion in the experimental conditions. PL spectra of ideal m-BN evaluated at different temperatures. Panel (b) PL spectra of m-BN evaluated at $T=30$ K using the scissor approximation for the band structure. The theoretical spectra have been blueshifted by 0.67~eV in order to match the QP peak of the experimental spectrum from Ref.\cite{doi:10.1021/acs.nanolett.1c02531}. In addition all the spectra have been renormalized to the value of corresponding QP peak.}\label{fig4}
\end{figure}

Interestingly, despite the strong el-ph interaction arising from the long range Fr\"ohlich contribution\cite{PhysRevB.105.115414} ($|g|^2$ $\sim$ 4 eV$^2$ at $\Gamma$), the strength of the exc-ph coupling for the LO phonon is smaller than in the other phonon channels (Fig. \ref{fig2} (e)). This suggests that localization associated to the electron-hole correlation plays a key role in suppressing long range effects in the coupling with lattice vibrations. This effect arises from the intrinsic structure of the exc-ph matrix element which consists of two opposite contributions ($\mathcal{A}$ and $\mathcal{B}$) that tend to cancel each other. In particular, for $\bar{\bfq}\rightarrow 0$, the cancellation is exact in the limiting case in which the electron-phonon matrix elements for valence and conduction states are the same for each $\bfk$ point. This is just what happens for the Fr\"ohlich contribution to the el-ph matrix elements since it is independent from both the band index and the electron wave-vector\cite{PhysRevB.105.115414}. Of course the cancellation becomes partial at $\bar{\bfq}\neq 0$ and the exc-ph coupling takes a finite value that goes to zero at large $\bar{\bfq}$ consistently with the behaviour of the Fr\"ohlich interaction.

Cancellation effects, on the other hand, are negligible for the TO and ZO phonon modes at $\bar{\bfq}=K$. This can be qualitatively understood comparing the $\bfk$ distribution of the excitonic amplitude in Fig. \ref{fig1}(c)-(d) with the $\bfk$ distribution of the el-ph matrix elements shown in Fig. \ref{fig3}. First of all, we note that excitonic amplitude for the excitons $E^{(1)}_K$ and $E^{(2)}_K$ are localized on the $K'$ and and $K$ valleys, respectively. This ensure strong overlap with the $E^{(1)}_{\Gamma}$ excitons in the same region where the el-ph matrix elements are different from zero. Moreover, comparing Fig. \ref{fig3} (a) and (b), we clearly see that the strength of el-ph matrix elements to the valence states is negligible with respect to that to the conduction states. Thus, for the ZO phonon mode the exc-ph coupling is dominated by the $\mathcal{B}$ term. The opposite situation occurs for the TO phonon mode, where the $\mathcal{B}$ term is negligible, being the valence states the most coupled ones (compare Fig. \ref{fig3} (c) and (d)).

To conclude our study of the optical properties of m-BN, we analyze the effect of the exc-ph coupling on the emission spectra. From Fig. \ref{fig4} (c) we clearly see that the PL signal presents a double peak structure. The higher energy feature is associated to the direct recombination of the $E^{(1)}_{\Gamma}$ exciton while the other one is a satellite associated to the emission of a ZO phonon through the $E^{(1)}_{\Gamma}\leftrightarrow E^{(1)}_K$ scattering processes. As the temperature is lowered the occupation number of the $E^{(1)}_{\Gamma}$ exciton decreases and the direct peak is progressively suppressed. In particular, below $T=60$ K only the satellite contributes to the PL spectrum as expected for indirect gap semiconductors. Thus, the presence of satellites below the QP peak is tightly related to the presence of the $E^{(1)}_K$ valley and hence to the indirect band gap nature of the m-BN. As a matter of fact, the intra-valley coupling with the LO mode (i.e. the only one that would be present if the band gap was direct) is too weak to induce visible effects in the PL spectra. This is a consequence of the cancellation effects induced by electron-hole correlation and the fact that in 2D the Fr\"ohlich contribution to the el-ph interaction does not diverge for $\bfq\rightarrow 0$, as expected in 3D polar crystals,  but takes a finite value\cite{PhysRevB.105.115414,doi:10.1021/acs.nanolett.7b01090}.
 
At this point it is important to note that recent first-principles MBPT calculations suggested that the interaction with the substrate, always present in real experiments, pushes the nearly free-electron band at $\Gamma$ above the bottom of the conduction band at $K$, making m-BN a direct gap semiconductor\cite{PhysRevMaterials.7.024006}. Under these conditions, phonon satellites below the QP peak should be essentially absent in contrast with experimental observations\cite{Elias2019,doi:10.1021/acs.nanolett.1c02531}. Based on our analysis of the exc-ph coupling, a possible explanation for the phonon satellites observed in the experimental spectra is that m-BN is actually a multi-valley semiconductor consisting of two nearly degenerate exciton valleys namely: $E^{(1)}_{\Gamma}$ and $E^{(2)}_K$.
In this scenario, the lowering of $E^{(2)}_K$ valley could be caused by intrinsic effects such as many-body corrections to the static BSE that could modify the excitonic band structure as well as extrinsic effects related to the interaction with the substrate as for example the induced strain.~Assuming this is the case, satellites originate from inter-valley $E^{(1)}_{\Gamma}\leftrightarrow E^{(2)}_K$ scattering processes mediated by the TO phonon mode. 

To investigate this possibility, we alternatively evaluated the PL spectrum using the Kohn-Sham eigenvalues with a rigid scissor of 2.8 eV in order to simulate the band gap opening at $K$. This ensures that the resulting QP electronic band structure is characterized by a direct band gap as expected in presence of the substrate\cite{PhysRevMaterials.7.024006}. In this way the $E^{(1)}_K$ exciton is pushed above the $E^{(2)}_K$ exciton, as schematically illustrated in Fig.\ref{fig4} (a) and (b). In addition, we artificially down-shifted the $E^{(2)}_{K}$ exciton by 0.17 eV in order to simulate a double valley configuration (see blue line in Fig.\ref{fig4} (b)). In this way both excitons, $E^{(1)}_{\Gamma}$ and $E^{(2)}_K$ are populated. The resulting spectrum [green line in Fig. \ref{fig4} (d)] presents, beside the QP peak, additional structures related to the TO phonon satellite and its replicas, matching the basic features observed in recent PL spectra measured using SiO$_2$ as substrate\cite{doi:10.1021/acs.nanolett.1c02531} [see red line in Fig. \ref{fig4} (d)]. However the theory is not able to capture the larger broadening observed in the experiment. This suggests that the interaction with the substrate may activate new channels for the exc-ph coupling involving lower energy acoustic modes that in the ideal m-BN are not coupled. The agreement with the experiment can be further improved applying an artificial broadening of 20 meV in order to simulate the effect of the acoustic phonons background (see blue line in Fig. \ref{fig4} (d)), although some features close to quasi-particle peak are still missing in theory. Their understanding requires a detailed study of the effect of the substrate on the exciton band structure and exc-ph coupling that goes beyond the scope of the present work.

We acknowledge fruitful discussion with G. Cassabois.
We acknowledge EuroHPC access on LUMI (EHPC-REG-2022R03-090) for high performance computing resources. This work was funded by the European Union (ERC, DELIGHT, 101052708). Views and opinions expressed are however those of the author(s) only and do not necessarily reflect those of the European Union or the European Research Council. Neither the European Union nor the granting authority can be held responsible for them.

\bibliography{bibliography}

\begin{mcitethebibliography}{46}
\bibitem[lot()]{loto}
For simplicity and with abuse of language, we use the ZO,LO,TO terminology also
  outside of $\mathbf{q} = \Gamma$\relax
\mciteBstWouldAddEndPuncttrue
\mciteSetBstMidEndSepPunct{\mcitedefaultmidpunct}
{\mcitedefaultendpunct}{\mcitedefaultseppunct}\relax
\EndOfBibitem
\bibitem[foo()]{footnote}
We emphasize that the negiglible linewidth is due to the energy of the
  $E^{(1)}_K$ exciton and hence to the relative position of the conduction band
  at $\Gamma$ with respect to the top of the valence band at $K$. Indeed, even
  a small decrease in the energy of the $E^{(1)}_K$ state induces a sizeable
  red shift of the onset of the exciton self-energy resulting in a remarkable
  enhancement of the exciton linewidth and a red shift of the ZO
  satellite.\relax
\mciteBstWouldAddEndPunctfalse
\mciteSetBstMidEndSepPunct{\mcitedefaultmidpunct}
{}{\mcitedefaultseppunct}\relax
\EndOfBibitem
\end{mcitethebibliography}

\providecommand{\latin}[1]{#1}
\makeatletter
\providecommand{\doi}
  {\begingroup\let\do\@makeother\dospecials
  \catcode`\{=1 \catcode`\}=2 \doi@aux}
\providecommand{\doi@aux}[1]{\endgroup\texttt{#1}}
\makeatother
\providecommand*\mcitethebibliography{\thebibliography}
\csname @ifundefined\endcsname{endmcitethebibliography}
  {\let\endmcitethebibliography\endthebibliography}{}

\end{document}